\documentstyle[aas2pp4]{article}

\slugcomment{Not to appear in Nonlearned J., 45.}

\lefthead{}
\righthead{}

\begin{document}

\title{How Dim Accreting Black Holes Could Be?}

\author{Marek A. Abramowicz\altaffilmark{1} and Igor V.
Igumenshchev\altaffilmark{2,3}}
\affil{Harvard-Smithsonian Center for Astrophysics, 60 Garden Street,
Cambridge MA 02138}

\altaffiltext{1}{Department of Astronomy \& Astrophysics, G\"oteborg
University \& Chalmers University of Technology, 412-96 G\"oteborg,
Sweden}

\altaffiltext{2}{Laboratory for Laser Energetics, University of
Rochester, 250 East River Road, Rochester NY 14623-1299.} 

\altaffiltext{3}{Institute of Astronomy,
48 Pyatnitskaya Ulitsa, 109017 Moscow, Russia.}

\begin{abstract}

Recent hydrodynamical simulations of radiatively inefficient black hole
accretion flows with low viscosity have demonstrated that these flows differ
significantly from those described by an advection-dominated model.
The black hole flows are advection-dominated only in their inner parts,
but $convectively$ dominated at radii $R \ga 10^2 R_G$. In such flows, the
radiative output comes mostly from the convection part, and the radiative
efficiency is independent of accretion rate and equals $\epsilon_{BH}
\simeq 10^{-3}$. This value gives a limit for how dim an accreting
black hole could be. It agrees with recent $Chandra$ observations which
indicate that accreting black holes in low-mass X-ray binaries
are by factor about 100
dimmer than neutron stars accreting with the same accretion rates.

\end{abstract}

\keywords{accretion, accretion disks --- black hole physics --- convection}

\section{Introduction}

It was recognized already a long time ago that physics of the innermost
parts of radiatively inefficient
black hole accretion disks is dominated by advection 
(e.g. Abramowicz 1981; Gilham 1981; Begelman \& Meier 1982).  
Advection-dominated accretion flow (ADAF)  is a simple analytic model of
such flows, constructed under the assumption that advection cooling
dominates not only close to the black hole, but at all radii 
(Narayan \& Yi 1994; Abramowicz et al. 1995). ADAFs provided a
remarkably accurate $quantitative$ prediction of detailed shapes of
electromagnetic spectra of observed accreting black holes (see reviews in
Narayan, Mahadevan \& Quataert 1998;  Kato, Fukue \& Mineshige 1998).  
Perhaps, the most interesting prediction of the ADAF model, based
$directly$ on the existence of the black hole event horizon, is that for
the same accretion rates, accreting black holes in compact binary systems
should be considerably dimmer, than accreting neutron stars (Narayan \& Yi
1995; Narayan, McClintock \& Yi 1996; Narayan, Barret \& McClintock 1997).
That is simply because the advected energy is lost inside the event
horizon of a black hole, while it is re-radiated from the surface of a
neutron star. This prediction was recently confirmed by the $Chandra$ data
on soft X-ray transients in a low-luminosity quiescent state (Menou et al.
1999; Garcia et al. 2001). Specifically, it was found that accreting black
holes are dimmer by the factor $f \simeq 100$ than neutron stars accreting
at the same rates.

In this $Letter$ we point out that the observed value of $f \simeq 100$
follows also from numerical models of radiatively inefficient
accretion flows that have been recently constructed.

\section{CDAFs versus ADAFs}

Self-similar ADAFs are convectively unstable, as was demonstrated by
Gilham (1981), Begelman \& Meier (1982) and Narayan \& Yi (1994). It was
firmly established in several numerical works (Igumenshchev, Chen \&
Abramowicz 1996; Igumenshchev \& Abramowicz 1999, 2000, 2001; Stone,
Pringle \& Begelman 1999) that the occurrence of this instability was not
an artifact of simplifying assumptions adopted in self-similar ADAF
models, but instead it is a genuine physical property of such flows with
low viscosity (viscosity coefficient $\alpha\la 0.1$).  Note that the
values $\alpha \la 0.1$ is consistent with the present understanding of
the origin of turbulent viscosity in accretion flows -- it follows from
simulations of the Balbus-Hawley instability (e.g. review in Hawley \&
Balbus 1995).
Our recent numerical modeling of magneto-hydrodynamical (MHD) radiatively
inefficient accretion flows (to be published) have shown that convection
motions may develop
in such flows.
Results of these simulations clearly demonstrate close qualitative
analogy between
MHD and low viscosity hydrodynamical simulations of accretion flows.
In subsequent publications we shall discuss in details our MHD
simulations,
and in particular explain why results of any realistic MHD simulations
$should$ differ from those
recently reported by Hawley, Balbus \& Stone (2001).

Thus, according to present understanding of turbulent viscosity in
accretion flows and estimates of its magnitude, radiatively inefficient
accretion flows $must$ $be$ convective. Convection is not a small
perturbation to ADAFs, instead it significantly changes the radial
structure of such flows.  With respect to ADAFs (with $\rho \propto
R^{-3/2}$), the convection-dominated accretion flows (CDAFs) have a much
flatter radial density profile, $\rho \propto R^{-1/2}$, and reduced mass
accretion rates. Convection carries angular momentum radially inwards and
energy outwards.  Simple analytic self-similar models of CDAFs which
reproduce all these properties and reasonably agree with numerical models
have been constructed by Narayan, Igumenshchev \& Abramowicz (2000) and
Quataert \& Gruzinov (2000). An analysis of self-similar CDAFs have shown
that the radiatively inefficient accretion flows are advection-dominated
in the inner part ($R\la 100 R_G$) and convection-dominated in the outer
part (Abramowicz et al. 2001). However, the presence of
the relatively small inner advection-dominated part in such flows makes
very little change to their global structure well described by the CDAF
model.

For the present discussion of how dim accreting black holes could be, it
is important to note that CDAFs have a significant (with respect to ADAFs)
outward energy flux carried by convection, with the ``luminosity''
$L_c=\epsilon_c \dot{M}c^2$, where $\dot{M}$ is the mass accretion rate,
and $\epsilon_c$ is the ``efficiency''. From numerically constructed
hydrodynamical models one concludes that 
$\epsilon_c\approx 3\times 10^{-3}-10^{-2}$ and weakly
depends on parameters of the problem, which are the adiabatic index
$\gamma$ and the viscosity parameter $\alpha$. The value of $\epsilon_c$
is independent of the accretion rate. CDAFs radiate mostly at their outer
parts due to the flattened density profile (Igumenshchev 2000;
Igumenshchev \& Abramowicz 2000). Ball, Narayan \& Quataert (2001) have
studied radiative processes that will transform a fraction $\eta$ of the
outward convection energy flux into radiation. They have concluded that
the most important radiative process is bremsstrahlung and most of the
energy is radiated in X-rays.  They pointed out that although at present
one could give only a very rough estimate for $\eta$, because of purely
known conditions in accretion flows on outside, it is unlikely that $\eta$
is significantly smaller than 1.  In this $Letter$ we assume that
$\eta\simeq 0.1-1$.

We conclude, based on the given above arguments, that the radiative output
of radiatively inefficient black hole accretion flows is determined by the
convection part of the flow. It peaks at X-ray frequencies and has the
radiative efficiency $\epsilon_{BH} = \eta\epsilon_c\simeq 10^{-3}$.

Note that our estimate of $\epsilon_{BH}$ is based on the assumption of
negligible heating of electrons during the viscous dissipation of
rotational and gravitational energies in accretion flows. If electrons are
heated more significantly through magnetic reconnections, as
Bisnovatyi-Kogan \& Lovelace (1997) argued, the value of $\epsilon_{BH}$
could be much larger.
However, there is no robust theoretical estimates of
importance of this effect (Quataert \& Gruzinov 1999).
 
\section{Discussion and conclusion}

Narayan's prediction that the event horizon makes radiatively inefficient
black hole flows much dimmer that neutron star flows at the same accretion
rates was confirmed by the $Chandra$ data (Menou et al. 1999; Garcia et
al. 2001).  These authors (and Lasota 2000) have convincingly argued that
the observed X-ray luminosities come from accretion power, and therefore
the observed difference -- black holes are about 100 times dimmer than
neutron stars -- should be explained by difference in radiative efficiency
of accretion flows.
We point out that for neutron stars the efficiency has its standard value
$\epsilon_{NS} \approx 10^{-1}$ (e.g. Frank, King \& Raine 1992), 
and therefore the factorial difference in
luminosities $f = \epsilon_{NS}/\epsilon_{BH} \simeq 100$ as indeed
observed.

\acknowledgments
The work was supported by NSF grant PHY 9507695, RFBR grant 00-02-16135,
and Swedish NFR grant.

\appendix

\clearpage

%
%

\end{document}